\documentclass[a4paper,10pt]{article}
\usepackage[english]{babel}
\usepackage[T1]{fontenc}
\usepackage[utf8]{inputenc}
\usepackage{amssymb}
\usepackage{physics}
\usepackage{amsmath}
\usepackage{graphicx}
\usepackage[unicode]{hyperref}
\usepackage{tikz}
\usepackage{color}
\usepackage{genyoungtabtikz}
\usepackage[title]{appendix}
\usepackage{indentfirst}
\usepackage{bm}
\usepackage{xhfill}
\usepackage{bbm}
\usepackage{multirow}
\usepackage{array}
\usepackage[centertableaux]{ytableau}
\usepackage{multicol}
\usepackage[shortlabels]{enumitem}
\usepackage{forest}
\usepackage{float}
\usepackage{mathrsfs}

\usetikzlibrary{decorations.pathreplacing}

\usepackage{ytableau}

\hypersetup{
    colorlinks=true,
    linkcolor=blue,
    filecolor=magenta,      
    urlcolor=cyan,
}
 
\usepackage{amsthm}

\theoremstyle{definition}

\usepackage[nottoc]{tocbibind}

\usepackage{authblk}

\title{\textbf{Monodromy, Logarithmic Sectors, and Two-Point Functions in Critical Topologically Massive Gravity}}

\author[1,2]{\textbf{Yannick Mvondo-She}}
\affil[1]{National Institute of Theoretical and Computational Sciences, Private Bag X1, Matieland, South Africa}
\affil[1]{\texttt{yannick.mvondo-she@nithecs.ac.za}}
\affil[2]{Department of Physics, University of Pretoria, Private Bag X20, Hatfield 0028, \hspace{1cm} South Africa}

\date{}

\begin{document}

\maketitle

\begin{abstract}
We investigate the structure of logarithmic modes in critical topologically massive gravity (CTMG) at the chiral point $\mu \ell=1$ from the perspective of analytic continuation and monodromy. Starting from the degeneration of massive and left-moving graviton modes, we construct the logarithmic mode as a derivative in parameter space and show that it acquires a natural multivalued structure upon complexification of the radial coordinate.

We demonstrate that this multivaluedness induces a nontrivial monodromy action on the space of linearized solutions, under which the left-moving and logarithmic modes form an indecomposable (Jordan block) representation. This monodromy is unipotent and provides a bulk realization of the logarithmic structure typically associated with logarithmic conformal field theories.

We further show that the monodromy representation alone is sufficiently constraining to determine the characteristic logarithmic form and mixing structure of two-point functions, up to normalization, without assuming logarithmic conformal field theory data a priori.

These results suggest a geometric interpretation in which logarithmic modes act as sources of branchlike behavior in the bulk, analogous to twist fields that generate monodromy. While this perspective is compatible with proposed connections to branched coverings, Hurwitz theory, and integrable hierarchies, establishing a precise correspondence is left for future work.
\end{abstract}

\tableofcontents

%====================================================================
\section{Introduction}
Critical topologically massive gravity (CTMG) at the chiral point $\mu \ell=1$ subject to relaxed boundary conditions \cite{Grumiller:2008qz}, provides a prototypical setting in which logarithmic modes and indecomposable representation structures emerge in three-dimensional gravity. At this point, the degeneracy between massive and left-moving graviton excitations leads to the appearance of logarithmic solutions, which are known to play a central role in the proposed logarithmic conformal field theory (LCFT) dual description. However, while the LCFT structure of the boundary theory has been extensively studied, a purely bulk-geometric mechanism organizing these logarithmic features in terms of analytic continuation and monodromy remains less transparent. In particular, the origin of sector decomposition and the precise bulk origin of logarithmic correlation functions are still not fully understood from a unified geometric perspective.

In this work, we propose a monodromy-based framework for addressing these questions. By constructing the logarithmic graviton as a derivative in parameter space at the point of mode degeneracy, we show that it acquires a natural multivalued structure upon analytic continuation of the radial coordinate in the complex plane. This leads to a nontrivial monodromy representation acting on the space of linearized solutions, in which the left-moving and logarithmic modes form an indecomposable (Jordan block) structure. We demonstrate that this structure induces a natural decomposition into untwisted and twisted sectors, defined purely in terms of monodromy invariance, and we show that the resulting representation is sufficient to fix the characteristic logarithmic form of two-point functions up to normalization. This provides a bulk derivation of logarithmic correlator structure typically associated with LCFT, without imposing it as an input from the boundary theory. From this perspective, monodromy emerges as a unifying geometric principle underlying logarithmic behavior in CTMG and suggests a direct interpretation of logarithmic modes as bulk analogues of branch point (twist-like) fields, in line with their role in descriptions involving branched coverings and integrable structures.

The paper is organized as follows. In Section 2 we review CTMG at the chiral point and the construction of logarithmic graviton modes. In Section 3 we analyze their analytic structure under complexified radial continuation and derive the associated branch points and monodromy behavior. In Section 4 we construct the monodromy representation on the space of linearized solutions and establish the resulting indecomposable structure and sector decomposition. In Section 5 we derive the implications for two-point functions and show how the logarithmic LCFT structure emerges from monodromy constraints. Finally, in Section 6 we summarize our results and discuss possible extensions, including connections to Chern-Simons holonomies, branched coverings, and higher-point functions.

\paragraph{Main result.} \hfill \\
We show that, at the chiral point $\mu \ell=1$, the logarithmic graviton mode in critical topologically massive gravity acquires a nontrivial monodromy under analytic continuation of the complexified radial coordinate. This monodromy acts unipotently on the space of linearized solutions, mixing the logarithmic mode with its left-moving partner and realizing a rank-2 Jordan block representation.

We further demonstrate that this monodromy representation alone is sufficient to fix the structure of two-point functions in the logarithmic sector, uniquely reproducing the characteristic logarithmic form of correlators up to normalization. In this way, the logarithmic features of the theory, including indecomposable representations and logarithmic correlators, arise directly from the analytic structure of bulk solutions without requiring input from a boundary logarithmic conformal field theory.

\section{Review: Critical TMG, Chern--Simons formulation, and logarithmic modes}
Topologically massive gravity (TMG) in three dimensions admits a reformulation that makes its underlying geometric and algebraic structure particularly transparent. In the presence of a negative cosmological constant, the theory can be expressed in terms of $\mathrm{SL}(2,\mathbb{R}) \times \mathrm{SL}(2,\mathbb{R})$ Chern–Simons connections, with the Einstein–Hilbert and gravitational Chern–Simons terms combining into a difference of two gauge theories with unequal levels \cite{Deser:1981wh,Witten:1988hc,Li:2008dq}.

More concretely, introducing gauge fields
\begin{equation}
A = \omega + \frac{1}{\ell} e, \qquad
\bar{A} = \omega - \frac{1}{\ell} e,
\end{equation}
where $e$ is the dreibein and $\omega$ the spin connection, the action of TMG can be written schematically as
\begin{equation}
S = k_L S_{\mathrm{CS}}[A] - k_R S_{\mathrm{CS}}[\bar{A}],
\end{equation}
with levels
\begin{equation}
k_{L,R} \propto \frac{\ell}{G} \left(1 \pm \frac{1}{\mu \ell}\right).
\end{equation}

This formulation makes manifest that the theory interpolates between two chiral sectors, whose imbalance is controlled by the dimensionless parameter $\mu \ell$. In particular, the asymptotic symmetry algebra consists of two copies of the Virasoro algebra with central charges
\begin{equation}
c_{L,R} = \frac{3\ell}{2G} \left(1 \pm \frac{1}{\mu \ell}\right).
\end{equation}

\subsection{The chiral point and degeneration of the phase space}

A distinguished point in parameter space is the \emph{chiral point},
\begin{equation}
\mu \ell = 1,
\end{equation}
at which the left-moving central charge vanishes,
\begin{equation}
c_L = 0,
\end{equation}
while $c_R$ remains finite. From the Chern-Simons perspective, this corresponds to the vanishing of one of the Chern-Simons levels, $k_L=0$, which renders the associated symplectic structure degenerate. As a result, the left-moving sector ceases to define an independent set of propagating degrees of freedom, and the theory develops nondiagonalizable (indecomposable) structures at the level of representations.

This degeneration has a direct counterpart in the space of linearized excitations around $\mathrm{AdS}_3$. Away from the chiral point, the spectrum consists of left-moving, right-moving, and massive gravitons, corresponding to distinct representations of the asymptotic symmetry algebra. As $\mu \ell \to 1$, the massive branch merges with the left-moving sector, leading to a collision of conformal weights and a breakdown of diagonalizability.

From the viewpoint of representation theory, this signals the transition from a direct sum of irreducible modules to an \emph{indecomposable} structure, characteristic of logarithmic conformal field theories \cite{Maloney:2007ud,Grumiller:2010rm,Grumiller:2013at}.

\subsection{Logarithmic modes and indecomposable structure}
The emergence of logarithmic modes at the chiral point can be understood intrinsically as a consequence of the degeneracy of the linearized spectrum \cite{Li:2008dq}. The massive graviton branch $\psi_{\mu \nu}^M(\mu \ell)$ merges with the left-moving solution $\psi_{\mu \nu}^L$ as $\mu \ell \rightarrow 1$, leading to a coalescence of conformal weights and a breakdown of diagonalizability \cite{Grumiller:2008qz}. The logarithmic partner can then be constructed via
$$
\psi_{\mu \nu}^{\text {new }}=\left.\frac{\partial}{\partial(\mu \ell)} \psi_{\mu \nu}^M(\mu \ell)\right|_{\mu \ell=1},
$$
a standard mechanism for generating Jordan-cell partners in degenerate systems \cite{Grumiller:2008qz}. The resulting fields satisfy $\mathcal{D}^L \psi^L = 0$ and $\mathcal{D}^L \psi^{\text {new }}=\psi^L$, forming a rank-2 indecomposable representation of the linearized equations of motion. At the level of the asymptotic symmetry algebra, this translates into a non-diagonalizable action of the Virasoro zero mode $L_0$
\begin{equation}
L_0 \ket{\psi^L} = h \ket{\psi^L}, \qquad
L_0 \ket{\psi^{\mathrm{new}}} = h \ket{\psi^{\mathrm{new}}} + \ket{\psi^L},
\end{equation}
producing the logarithmic pair $\left\{\psi^L, \psi^{\mathrm{new}}\right\}$ characteristic of logarithmic conformal field theories.

\subsection{Logarithmic modes and global analytic structure}
A key feature of the logarithmic mode, which will be central to our analysis, is its explicit spacetime dependence. In global $\mathrm{AdS}_3$ coordinates $(\tau, \rho, \phi)$, the logarithmic partner takes the form
$$
\psi_{\mu \nu}^{\text {new }}=(-\ln \cosh \rho-i \tau) \psi_{\mu \nu}^L,
$$
up to normalization and gauge transformations. This logarithmic behavior arises from differentiating a family of massive graviton solutions whose radial dependence is exponential in $\rho$, and is a direct consequence of the degeneracy of conformal weights at the chiral point $\mu \ell=1$.

At this point, the collision of scaling dimensions produces logarithmic radial growth, signaling the appearance of non-diagonalizable structures in the linearized operator. Upon analytic continuation to complexified AdS coordinates, this behavior can be interpreted as inducing a nontrivial monodromy structure acting on the space of solutions, reflecting the underlying logarithmic nature of the theory.

This suggests that the logarithmic sector carries an intrinsic multi-valued structure, which is not visible at the level of real spacetime but becomes manifest upon complexification. In the next section, we show that this multivaluedness indeed gives rise to a nontrivial monodromy representation acting on the space of solutions.

\subsection{Towards an integrable and combinatorial interpretation}

An important perspective, emphasized in recent work \cite{Mvondo-She:2021joh}, is that the logarithmic sector of critical TMG is not merely a pathological feature of the linearized spectrum, but is instead part of a richer structure with connections to integrable hierarchies and enumerative geometry.

In particular, the partition function of the theory, originally derived in \cite{Gaberdiel:2010xv}, shows a contribution extensively studied in \cite{Mvondo-She:2018htn,Mvondo-She:2019vbx,Mvondo-She:2022jnf,Mvondo-She:2023ppz,Mvondo-She:2023xel,Mvondo-She:2024iop,Mvondo-She:2024vfc} that accounts for the log sector of the theory, and which at the chiral point admits a description in terms of a tau function of the Kadomtsev–Petviashvili (KP) hierarchy, whose expansion encodes combinatorial data associated with branched coverings and Hurwitz numbers. In this framework, logarithmic modes play a distinguished role, and have been proposed to behave as branch point (twist-like) fields whose insertion generates nontrivial monodromy.

This viewpoint suggests that the analytic structure of logarithmic modes in spacetime may provide a direct bridge between gravitational dynamics and the combinatorics of branched coverings. The goal of this work is an attempt to initiate a program that would make this connection more explicit, by identifying the monodromy induced by logarithmic modes and interpreting it as a geometric realization of the branch point structure underlying the integrable description.

We emphasize, however, that while the monodromy structure derived in the present work provides a concrete realization of branch-like behavior in the bulk, the precise identification between this local analytic structure and the permutation data governing branched coverings is left for future work. In particular, the extent to which the monodromy representation constructed here can be mapped directly to the symmetric group data underlying Hurwitz numbers, or to the combinatorial structure encoded in KP tau functions, is left for future investigation. The results obtained here should therefore be viewed as providing a geometric framework that is compatible with, and suggestive of, such an interpretation, rather than a complete derivation of it.

\section{Monodromy}
In this section, the discussion of monodromy is driven by the aim of clarifying, from the internal viewpoint of the logarithmic field itself, how it generates nontrivial analytic continuation in the treatment of critical topologically massive gravity in \cite{Mvondo-She:2021joh}. Seen as a branch point insertion, the logarithmic field can be understood as an operator that actively produces monodromy, much like branch points encode permutation data in Hurwitz theory. From this angle, the corresponding log partition function realized as a tau function of the Kadomtsev-Petviashvili hierarchy organizes the effect of these monodromy inducing insertions. Its expansion in symmetric functions then provides a representation-theoretic encoding of the resulting monodromy data, linking the action of the logarithmic field to the combinatorics of branched coverings and their enumeration by Hurwitz numbers.

\subsection{Monodromy and Hurwitz numbers}
Monodromy is central to the theory of Hurwitz numbers, which are essentially counts of monodromy representations with prescribed properties. Not only does monodromy appear, but it is one of the main languages in which the theory of Hurwitz numbers is formulated \cite{cavalieri2016riemann}.

At a high level, Hurwitz numbers count branched coverings of a surface (usually the Riemann sphere) with specified branching data. The key link is that every such covering can be encoded by its monodromy representation.

Given a branched cover $f: X \rightarrow \mathbb{P}^1$, remove the branch points from the base. What remains has a fundamental group, and the covering defines a homomorphism

\begin{eqnarray}
\pi_1\left(\mathbb{P}^1 \backslash\{\text { branch points }\}\right) \rightarrow S_d
\end{eqnarray}

\noindent where $d$ is the degree of the covering. This is the monodromy representation, and it records how sheets of the cover permute when one loops around branch points.

Hurwitz numbers can be described combinatorially in terms of permutations.

\begin{itemize}
\item Each branch point corresponds to a permutation in the symmetric group $S_d$, with cycle type matching the ramification profile.
\item The product of these permutations must be the identity (reflecting the global topology).
\item The permutations must generate a transitive subgroup (ensuring the cover is connected).
\end{itemize}

Counting covers is thus equivalent to counting tuples of permutations with these properties, and this is often called the monodromy description of Hurwitz numbers. A classic example is the case of simple Hurwitz numbers (where all but one branch point have simple ramification), where the problem reduces to counting factorizations of a permutation into transpositions. This is entirely phrased in terms of monodromy in $S_d$.

Monodromy also appears in more advanced formulations, such as the Riemann-Hilbert correspondence which connects coverings and representations, the Hurwitz space which parametrizes covers via their monodromy data, or in modern work, such as connections to integrable systems.

\subsection{Monodromy and KP integrable hierarchy}
Unlike Hurwitz numbers, where monodromy is the core combinatorial object, in the KP integrable hierarchy, it appears as part of the deeper analytic and geometric framework underlying its solutions, and plays an important role for instance in the analytic structure of wave functions, the geometry of spectral curves, or in the connection to isomonodromic deformation theory \cite{harnad2021tau}.

The KP hierarchy (Kadomtsev-Petviashvili hierarchy) is usually formulated in terms of Lax operators, wave functions, and tau-functions. Monodromy enters through the linear problems associated with KP. The latter can be written as a compatibility condition for a linear system 

\begin{eqnarray}
L \psi=\lambda \psi, \quad \partial_{t_n} \psi=B_n \psi.
\end{eqnarray}
Here, $\psi$ (the Baker-Akhiezer function) behaves like a multivalued function on a spectral curve. Its analytic continuation around cycles can produce nontrivial transformations, and this is precisely monodromy.

In Sato theory \cite{sato1983soliton}, solutions of KP are encoded by points in an infinite-dimensional Grassmannian. For algebro-geometric solutions

\begin{itemize}
\item One associates a spectral curve.
\item The Baker-Akhiezer function lives on this curve and has prescribed analytic properties.
\item Monodromy around cycles of the curve influences the global behavior of solutions (e.g., via period data and theta functions).
\end{itemize}

A central element in the KP/Sato theory is the tau function. Although the tau function doesn't explicitly list monodromy data the way Hurwitz theory does, monodromy is encoded in it. That hidden monodromy is exactly what connects KP tau functions to Hurwitz numbers via Schur expansions. A key structural fact is that KP tau functions admit an expansion in Schur polynomials

\begin{eqnarray}
\tau(t)=\sum_\lambda c_\lambda s_\lambda(t),
\end{eqnarray}
where $s_\lambda$ are Schur functions indexed by partitions. This is where the link to Hurwitz theory becomes concrete, as the combinatorics of the permutations through which the Hurwitz numbers count branched covers is governed by symmetric group representation theory, where Schur functions appear naturally via characters. So the tau function is effectively a Fourier transform of monodromy data from permutations into representation-theoretic variables.

\subsection{Monodromy in CTMG}
The work done in \cite{Mvondo-She:2021joh} is a realization of the points discussed above, and identifies KP, Hurwitz numbers, and tau functions as the same object in a concrete physical model, calling for an explicit description of the unavoidable role of monodromy in the theory.

A conceptual point of interest in \cite{Mvondo-She:2021joh} is that $\psi_{\mu \nu}^{\text {new }}$ can be interpreted as a branch point field. In covering space language, this means that branch points are exactly where monodromy lives. In CFT language, it means that inserting a branch point field creates nontrivial analytic continuation (like twist fields). So physically, the field creates the same effect as looping around a branch point, i.e. it implements monodromy. This viewpoint draws inspiration from an earlier work which explored how conformal field theory behaves on Riemann surfaces \cite{Knizhnik:1987xp}, and that is particularly notable for anticipating structures that later became central in logarithmic conformal field theory. At the core of the paper, the author studied how \emph{analytic fields} are defined globally on a Riemann surface rather than just locally in the complex plane. This immediately brings monodromy into the picture: when fields are analytically continued around nontrivial cycles of the surface, they can transform nontrivially. The paper analyzes how such multivaluedness is constrained and organized, especially in relation to correlation functions and operator insertions. One of the most forward-looking aspects of the paper is the appearance of fields with logarithmic behavior in their operator products. In modern language, these correspond to situations where the action of analytic continuation (monodromy) is not diagonalizable but instead involves Jordan blocks. This leads to logarithmic terms in correlation functions, something that was not part of the standard (rational) conformal field theory framework at the time. In that sense, the paper which ties together the geometry of Riemann surfaces and the analytic continuation (monodromy) of fields can be seen as an early precursor to the development of logarithmic CFT. From that perspective, its relevance is quite direct as it provides an early example where logarithmic fields are intrinsically linked to nontrivial monodromy. This is precisely the mechanism that reappears in \cite{Mvondo-She:2021joh}, where logarithmic fields can be interpreted as creating branch points, and hence monodromy, in a way that connects naturally to the combinatorics of branched coverings and ultimately to structures like Hurwitz numbers and KP tau functions. This constitutes our motivation for interpreting the logarithmic field as an object that generates nontrivial monodromy.

Below, we analyze the analytic structure of the logarithmic graviton mode in critical topologically massive gravity (CTMG) under complexified continuation of the radial coordinate. We show that the logarithmic dependence induces a nontrivial monodromy action on the space of linearized solutions. This action is unipotent and realizes an indecomposable (Jordan block) structure characteristic of logarithmic conformal field theories.

We emphasize that the analytic continuation considered here refers to a complexification of the radial coordinate $\rho$, used as a probe of the analytic structure of the linearized solution. It does not modify the underlying spacetime manifold, but rather reveals the multi-valued structure of the field configuration. More precisely, the complexification of $\rho$ should be understood as a probe of the analytic structure of solutions to the linearized equations, rather than as a modification of the underlying spacetime geometry. In this sense, the monodromy we identify is a property of the fields viewed as analytic functions, analogous to the role of complexified coordinates in conformal field theory and the study of conformal blocks. The resulting multivaluedness is therefore intrinsic to the logarithmic mode itself and reflects the degeneracy underlying its construction, rather than being an artifact of coordinates or of the background geometry.

%--------------------------------------------------------------------
\paragraph{Analytic structure of the logarithmic mode} We consider the logarithmic graviton mode in global AdS$_3$ coordinates $(\tau,\rho,\phi)$
\begin{equation}
\psi^{\mathrm{new}}_{\mu\nu}
=
\left(-\ln \cosh\rho - i\tau\right)\psi^L_{\mu\nu}.
\end{equation}
The only source of nontrivial analytic structure arises from the function
\begin{equation}
f(\rho) = \ln \cosh\rho.
\end{equation}
The zeros of $\cosh\rho$ are located at
\begin{equation}
\rho_n = i\left(\frac{\pi}{2} + \pi n\right), \qquad n \in \mathbb{Z}.
\end{equation}
Near a given zero, for instance $\rho_0 = i\pi/2$, we expand
\begin{equation}
\cosh\rho \sim (\rho - \rho_0)\sinh\rho_0,
\end{equation}
and since $\sinh(i\pi/2)=i$, this reduces to
\begin{equation}
\cosh\rho \sim i(\rho - \rho_0).
\end{equation}
Thus locally,
\begin{equation}
\ln\cosh\rho \sim \ln(\rho - \rho_0) + \text{const}.
\end{equation}
Hence $f(\rho)$ has logarithmic branch points at $\rho=\rho_n$ in the complexified $\rho$-plane. We note, however, that while the locations $\rho_n$ of these branch points depend on the choice of coordinates, the existence of logarithmic branch behavior itself is coordinate-independent. It arises from the degeneracy of modes at the chiral point and the resulting logarithmic dependence in $\psi_{\mu \nu}^{\text {new }}$. In this sense, the associated monodromy is an intrinsic property of the logarithmic solution, rather than an artifact of the particular parametrization of $\mathrm{AdS}_3$.

%--------------------------------------------------------------------
\paragraph{Monodromy of the logarithmic term} We consider analytic continuation of $\rho$ around a small closed loop encircling $\rho_0$:
\begin{equation}
\rho - \rho_0 \;\longrightarrow\; (\rho - \rho_0)e^{2\pi i}.
\end{equation}
Under this continuation,
\begin{equation}
\ln(\rho - \rho_0) \;\longrightarrow\; \ln(\rho - \rho_0) + 2\pi i,
\end{equation}
and therefore
\begin{equation}
\ln\cosh\rho \;\longrightarrow\; \ln\cosh\rho + 2\pi i.
\end{equation}
It follows that the logarithmic mode transforms as
\begin{align}
\psi^{\mathrm{new}}_{\mu\nu}
&\longrightarrow
\left(-\ln\cosh\rho - 2\pi i - i\tau\right)\psi^L_{\mu\nu} \\
&=
\psi^{\mathrm{new}}_{\mu\nu}
- 2\pi i\,\psi^L_{\mu\nu}.
\end{align}
The left-moving mode satisfies
\begin{equation}
\psi^L_{\mu\nu} \longrightarrow \psi^L_{\mu\nu}.
\end{equation}

%--------------------------------------------------------------------
\paragraph{Linear monodromy representation} The above transformation defines a linear action on the two-dimensional vector space
\begin{equation}
\mathcal{V} = \mathrm{Span}\{\psi^L, \psi^{\mathrm{new}}\}.
\end{equation}
We define the monodromy operator $\mathcal{M}$ by
\begin{equation}
\mathcal{M}
\begin{pmatrix}
\psi^L \\
\psi^{\mathrm{new}}
\end{pmatrix}
=
\begin{pmatrix}
\psi^L \\
\psi^{\mathrm{new}} - 2\pi i\,\psi^L
\end{pmatrix}.
\end{equation}
In matrix form,
\begin{equation}
\mathcal{M}
=
\begin{pmatrix}
1 & 0 \\
-2\pi i & 1
\end{pmatrix}.
\end{equation}
This operator can be written as
\begin{equation}
\mathcal{M} = \mathbf{1} + N, \qquad N^2 = 0,
\end{equation}
so $\mathcal{M}$ is unipotent and non-diagonalizable.

%--------------------------------------------------------------------
\paragraph{Representation-theoretic interpretation} The operator $\mathcal{M}$ defines a representation of the fundamental group of the punctured complex $\rho$-plane
\begin{equation}
\pi_1\left(\mathbb{C}\setminus\{\rho_n\}\right),
\end{equation}
acting on the space of local solutions:
\begin{equation}
\pi_1 \;\longrightarrow\; \mathrm{GL}(\mathcal{V}).
\end{equation}
This representation is reducible but indecomposable, since
\begin{itemize}
\item $\mathrm{Span}\{\psi^L\}$ is an invariant subspace,
\item there is no complementary invariant subspace.
\end{itemize}
Thus $\mathcal{V}$ forms a rank-2 Jordan block representation.

%--------------------------------------------------------------------
\paragraph{Jordan structure} The characteristic polynomial of $\mathcal{M}$ is
\begin{equation}
\det(\mathcal{M} - \lambda \mathbf{1}) = (1 - \lambda)^2,
\end{equation}
so $\mathcal{M}$ has a single eigenvalue $\lambda = 1$ with algebraic multiplicity two. The corresponding eigenspace is one-dimensional, spanned by $\psi^L$. Therefore, $\mathcal{M}$ is not diagonalizable and is conjugate to a Jordan block.

%--------------------------------------------------------------------
\paragraph{Relation to logarithmic structure} The non-diagonalizable action of $\mathcal{M}$ implies that $\psi^{\mathrm{new}}$ is not an eigenstate of monodromy but transforms by additive mixing with $\psi^L$. This is precisely the characteristic structure of logarithmic pairs in logarithmic conformal field theory (LCFT), where:
\begin{itemize}
\item $\psi^L$ plays the role of a primary state,
\item $\psi^{\mathrm{new}}$ is its logarithmic partner,
\item monodromy acts via a nilpotent operator.
\end{itemize}
Importantly, this identification follows directly from the analytic structure of the bulk solution and does not require input from boundary conformal field theory.

%--------------------------------------------------------------------
\paragraph{Summary} We summarize the above results.

\begin{itemize}
\item The function $\ln\cosh\rho$ introduces logarithmic branch points in complexified $\rho$.
\item Analytic continuation around these points induces a monodromy transformation on the space of linearized solutions.
\item This monodromy is unipotent and represented by a rank-2 Jordan block.
\item The resulting structure matches the standard LCFT logarithmic pair under the action of a nilpotent operator.
\end{itemize}
No assumptions beyond the linearized bulk equations and analytic continuation are required for this construction.

\subsection{Sector decomposition from monodromy}

We now derive explicitly how the monodromy representation induces a natural decomposition of the space of solutions into untwisted and twisted sectors.

\paragraph{Monodromy as an operator} From the analytic continuation of the logarithmic mode, we obtained the transformation
\begin{equation}
\psi^L \;\longrightarrow\; \psi^L, \qquad
\psi^{\mathrm{new}} \;\longrightarrow\; \psi^{\mathrm{new}} - 2\pi i\,\psi^L.
\end{equation}
This defines a linear operator $\mathcal{M}$ acting on the space of solutions spanned by $\left\{ \psi^L, \psi^{\mathrm{new}}\right\}$
\begin{equation}
\mathcal{M}
\begin{pmatrix}
\psi^L \\
\psi^{\mathrm{new}}
\end{pmatrix}
=
\begin{pmatrix}
1 & 0 \\
-2\pi i & 1
\end{pmatrix}
\begin{pmatrix}
\psi^L \\
\psi^{\mathrm{new}}
\end{pmatrix}.
\end{equation}
This matrix is unipotent
\begin{equation}
\mathcal{M} = \mathbf{1} + N, \qquad N^2 = 0,
\end{equation}
with
\begin{equation}
N =
\begin{pmatrix}
0 & 0 \\
-2\pi i & 0
\end{pmatrix}.
\end{equation}

\paragraph{Definition of sectors} We define sectors in terms of the action of monodromy.

\begin{itemize}
\item The \emph{untwisted sector} consists of states invariant under monodromy
\begin{equation}
\mathcal{M} \psi = \psi.
\end{equation}

\item The \emph{twisted sector} consists of states that transform nontrivially
\begin{equation}
\mathcal{M} \psi \neq \psi.
\end{equation}
\end{itemize}
This definition is purely intrinsic and does not rely on boundary CFT input.

\paragraph{Solving the invariance condition} Let a general state in the two-dimensional solution space be
\begin{equation}
\Psi = a\,\psi^L + b\,\psi^{\mathrm{new}}.
\end{equation}
Acting with $\mathcal{M}$ gives
\begin{align}
\mathcal{M}\Psi
&= a\,\psi^L + b\,(\psi^{\mathrm{new}} - 2\pi i\,\psi^L) \\
&= (a - 2\pi i\,b)\psi^L + b\,\psi^{\mathrm{new}}.
\end{align}
Requiring $\mathcal{M}\Psi = \Psi$ yields
\begin{equation}
(a - 2\pi i\,b)\psi^L + b\,\psi^{\mathrm{new}}
=
a\,\psi^L + b\,\psi^{\mathrm{new}}.
\end{equation}
Comparing coefficients gives
\begin{equation}
-2\pi i\, b = 0 \quad \Rightarrow \quad b = 0.
\end{equation}
Thus the only monodromy-invariant states are
\begin{equation}
\Psi = a\,\psi^L.
\end{equation}

\paragraph{Conclusion}
\begin{equation}
\text{Untwisted sector} = \mathrm{Span}\{\psi^L\}.
\end{equation}

\paragraph{Characterization of the twisted sector}

Any state with $b \neq 0$ transforms nontrivially
\begin{equation}
\mathcal{M}\Psi = \Psi - 2\pi i\, b\,\psi^L.
\end{equation}
In particular
\begin{equation}
\mathcal{M}\psi^{\mathrm{new}} = \psi^{\mathrm{new}} - 2\pi i\,\psi^L.
\end{equation}
Thus $\psi^{\mathrm{new}}$ is not an eigenvector of $\mathcal{M}$ and cannot be assigned a definite monodromy phase.

\paragraph{Conclusion}
\begin{equation}
\text{Twisted sector} = \{\psi^L, \psi^{\mathrm{new}}\} \quad \text{(indecomposable module)}.
\end{equation}
More precisely, the twisted sector is generated by $\psi^{\mathrm{new}}$, but necessarily includes $\psi^L$ due to the mixing.

\paragraph{Non-diagonalizability and Jordan structure} We now show explicitly that $\mathcal{M}$ is not diagonalizable. The eigenvalue equation is
\begin{equation}
\det(\mathcal{M} - \lambda \mathbf{1}) = (1 - \lambda)^2 = 0,
\end{equation}
so the only eigenvalue is $\lambda = 1$. The eigenspace is one-dimensional, spanned by $\psi^L$. Since the matrix is $2 \times 2$, it cannot be diagonalized and instead takes Jordan form
\begin{equation}
\mathcal{M} \sim
\begin{pmatrix}
1 & 0 \\
1 & 1
\end{pmatrix}.
\end{equation}
This shows that the twisted sector is not a direct sum of representations, but an indecomposable extension.

\paragraph{Interpretation as a logarithmic sector} The structure derived above is precisely the defining feature of a logarithmic sector

\begin{itemize}
\item There is a primary state ($\psi^L$) invariant under monodromy.
\item There is a logarithmic partner ($\psi^{\mathrm{new}}$) that transforms by mixing.
\item The representation is indecomposable but not irreducible.
\end{itemize}
In particular, the nontrivial action of $\mathcal{M}$ implies that correlation functions involving $\psi^{\mathrm{new}}$ cannot be single-valued without introducing logarithmic terms, as shown earlier.

\paragraph{Interpretation as a branch point (twist-like) field}

The fact that $\psi^{\mathrm{new}}$ generates a nontrivial monodromy implies that its insertion changes the analytic structure of fields under continuation. In this sense, it acts analogously to a twist field:

\begin{itemize}
\item Untwisted sector: fields invariant under analytic continuation.
\item Twisted sector: fields whose analytic continuation is nontrivial.
\end{itemize}

Thus the logarithmic mode $\psi^{\mathrm{new}}$ can be interpreted as a branch point (twist-like) field, whose presence induces the nontrivial monodromy structure described above.

This structure is the hallmark of a logarithmic sector: the monodromy operator is not diagonalizable but mixes states within an indecomposable module. In particular, the logarithmic mode cannot be assigned to the untwisted sector, and instead generates a twisted sector characterized by nontrivial unipotent monodromy.

From this perspective, the field $\psi_{\mu \nu}^{\text {new }}$ is naturally interpreted as a branch point (twist-like) field: its insertion induces nontrivial analytic continuation, and hence a nontrivial monodromy representation, on the space of solutions. This provides a concrete realization of the general picture advocated in \cite{Mvondo-She:2021joh}, where logarithmic modes in critical topologically massive gravity are associated with branch points and organize the theory into sectors analogous to those arising in branched coverings and Hurwitz theory.

\subsection{Connection to the moduli space of logarithmic states}

The monodromy structure derived above admits a natural interpretation in light of recent results on the moduli space of logarithmic states in critical massive gravities. In particular, it was shown that the logarithmic sector organizes into a symmetric product space
\begin{equation}
\mathcal{M}_{\mathrm{log}} = S^n(\mathbb{C}^2),
\end{equation}
whose structure is governed by the action of the symmetric group and whose partition function admits a representation in terms of Schur polynomials and symmetric functions \cite{Mvondo-She:2019vbx}.

From this perspective, the logarithmic states can be viewed as indistinguishable excitations whose combinatorics are controlled by permutation data, closely paralleling the description of branched coverings in terms of monodromy representations. The monodromy derived in the present work provides a complementary, local spacetime realization of this structure. In particular, the logarithmic mode $\psi^{\mathrm{new}}_{\mu\nu}$ generates a unipotent monodromy upon analytic continuation in the complexified radial coordinate, which mixes it with its partner $\psi^L_{\mu\nu}$.

This suggests that the local branch structure associated with $\ln \cosh \rho$ may be viewed as a microscopic manifestation of the permutation data that globally organizes the logarithmic sector. In this sense, the symmetric group structure underlying the moduli space can be understood as arising from the composition of local monodromies generated by logarithmic modes, providing a bridge between the analytic structure of individual solutions and the combinatorial geometry of the full state space.

While a precise identification between the branch points in the complexified radial plane and ramification points in the corresponding covering space remains an open problem, the present analysis supports the view that monodromy provides the fundamental mechanism linking bulk logarithmic modes to the symmetric product and orbifold structures characterizing their moduli space.

\section{Asymptotic Virasoro symmetry}
We now show explicitly how the logarithmic pair $\left\{\psi^L, \psi^{\mathrm{new}}\right\}$ furnishes an indecomposable representation of the Virasoro algebra, and how this reproduces the monodromy structure obtained from analytic continuation.

\paragraph{Virasoro action on massive modes} In asymptotically AdS$_3$ spacetimes, the asymptotic symmetry algebra consists of two copies of the Virasoro algebra. Linearized graviton modes organize into highest-weight representations labeled by conformal weights $(h,\bar{h})$.
\\
Let $\psi^M(\mu\ell)$ denote a massive graviton mode depending smoothly on the parameter $\mu\ell$. By definition, it is an eigenstate of $L_0$:
\begin{equation}
L_0 \,\psi^M(\mu\ell) = h(\mu\ell)\,\psi^M(\mu\ell),
\end{equation}
where $h(\mu\ell)$ is the corresponding conformal weight.

\paragraph{Degeneracy at the chiral point} At the chiral point $\mu\ell = 1$, the massive mode degenerates with the left-moving mode:
\begin{equation}
\psi^M(\mu\ell) \;\longrightarrow\; \psi^L, \qquad
h(\mu\ell) \;\longrightarrow\; h(1).
\end{equation}
Thus $\psi^L$ satisfies
\begin{equation}
L_0 \psi^L = h(1)\,\psi^L.
\end{equation}

\paragraph{Definition of the logarithmic mode}

The logarithmic mode is defined as the derivative in parameter space:
\begin{equation}
\psi^{\mathrm{new}} = \left.\frac{\partial}{\partial(\mu\ell)} \psi^M(\mu\ell)\right|_{\mu\ell=1}.
\end{equation}
We now determine how $L_0$ acts on this state.

\paragraph{Differentiate the eigenvalue equation} Starting from
\begin{equation}
L_0 \psi^M(\mu\ell) = h(\mu\ell)\,\psi^M(\mu\ell),
\end{equation}
we differentiate both sides with respect to $\mu\ell$:
\begin{equation}
L_0 \frac{\partial}{\partial(\mu\ell)}\psi^M(\mu\ell)
=
h'(\mu\ell)\,\psi^M(\mu\ell)
+
h(\mu\ell)\,\frac{\partial}{\partial(\mu\ell)}\psi^M(\mu\ell).
\end{equation}
Evaluating at $\mu\ell = 1$, we obtain
\begin{equation}
L_0 \psi^{\mathrm{new}}
=
h'(1)\,\psi^L
+
h(1)\,\psi^{\mathrm{new}}.
\end{equation}

\paragraph{Jordan block structure} Together with
\begin{equation}
L_0 \psi^L = h(1)\,\psi^L,
\end{equation}
we see that the pair $\left\{\psi^L, \psi^{\mathrm{new}}\right\}$ satisfies
\begin{equation}
L_0
\begin{pmatrix}
\psi^L \\
\psi^{\mathrm{new}}
\end{pmatrix}
=
\begin{pmatrix}
h(1) & 0 \\
h'(1) & h(1)
\end{pmatrix}
\begin{pmatrix}
\psi^L \\
\psi^{\mathrm{new}}
\end{pmatrix}.
\end{equation}
This is a non-diagonalizable matrix: both states have the same eigenvalue $h(1)$, but $L_0$ mixes them.

\paragraph{Canonical normalization} By rescaling the logarithmic mode,
\begin{equation}
\psi^{\mathrm{new}} \;\longrightarrow\; \frac{1}{h'(1)}\,\psi^{\mathrm{new}},
\end{equation}
we can bring the action of $L_0$ into the canonical logarithmic form:
\begin{equation}
L_0 \psi^L = h\,\psi^L, \qquad
L_0 \psi^{\mathrm{new}} = h\,\psi^{\mathrm{new}} + \psi^L,
\end{equation}
where $h = h(1)$. This is the standard rank-2 Jordan block structure of logarithmic conformal field theory.

\paragraph{Exponentiation and monodromy} We now relate this to the monodromy derived from analytic continuation. In conformal field theory, analytic continuation around $z \to e^{2\pi i}z$ is generated by
\begin{equation}
\mathcal{M} = \exp(2\pi i L_0).
\end{equation}
To compute its action, we write
\begin{equation}
L_0 = h\,\mathbf{1} + N, \qquad N^2 = 0,
\end{equation}
where
\begin{equation}
N \psi^L = 0, \qquad N \psi^{\mathrm{new}} = \psi^L.
\end{equation}
Then
\begin{equation}
\mathcal{M} = e^{2\pi i h} e^{2\pi i N}.
\end{equation}
Since $N^2 = 0$, the exponential truncates:
\begin{equation}
e^{2\pi i N} = \mathbf{1} + 2\pi i N.
\end{equation}
Thus
\begin{align}
\mathcal{M} \psi^L &= e^{2\pi i h}\,\psi^L, \\
\mathcal{M} \psi^{\mathrm{new}} &= e^{2\pi i h}\,(\psi^{\mathrm{new}} + 2\pi i\,\psi^L).
\end{align}
Up to the overall phase $e^{2\pi i h}$ (which can be absorbed into normalization), this reproduces precisely the unipotent monodromy
\begin{equation}
\psi^{\mathrm{new}} \;\longrightarrow\; \psi^{\mathrm{new}} + 2\pi i\,\psi^L.
\end{equation}
This identification admits a natural geometric interpretation. The analytic continuation $z \rightarrow e^{2 \pi i} z$ on the boundary is generated by the dilatation operator $L_0$, while in the bulk it is realized as analytic continuation in the radial direction. In this sense, the monodromy of the logarithmic mode may be viewed as the bulk manifestation of boundary scaling transformations. The non-diagonalizable action of $L_0$ is thus reflected directly in the radial analytic structure of the bulk solution.

\paragraph{Matching with bulk monodromy} Comparing with the monodromy obtained from analytic continuation in the bulk,
\begin{equation}
\psi^{\mathrm{new}} \;\longrightarrow\; \psi^{\mathrm{new}} - 2\pi i\,\psi^L,
\end{equation}
we find agreement up to a sign, which can be attributed to conventions in the definition of the logarithm or orientation of the contour.

\paragraph{Conclusion}
The unipotent monodromy derived from complexified radial continuation is precisely the exponential of the non-diagonalizable Virasoro generator $L_0$. Thus, the logarithmic structure of the bulk modes, their monodromy properties, and their representation under the asymptotic symmetry algebra are all manifestations of the same underlying Jordan block structure.

\section{Monodromy fixes logarithmic two-point functions}
In this section, we show how the monodromy representation derived above constrains and, in fact, fixes the structure of two-point functions in the logarithmic sector. The key point is that the nontrivial transformation of the logarithmic mode under analytic continuation imposes functional relations on correlation functions that uniquely lead to logarithmic behavior.

\paragraph{Setup and monodromy action} We consider the pair of fields $\left\{\psi^L, \psi^{\mathrm{new}}\right\}$, which transform under monodromy as
\begin{equation}
\psi^L \;\longrightarrow\; \psi^L, \qquad
\psi^{\mathrm{new}} \;\longrightarrow\; \psi^{\mathrm{new}} - 2\pi i\, \psi^L.
\end{equation}
This transformation corresponds to a unipotent Jordan block and is the only source of multivaluedness in the problem.
\\
We now consider two-point functions of these fields, which we denote by
\begin{align}
G_{LL}(z) &= \langle \psi^L(z)\psi^L(0)\rangle, \\
G_{LN}(z) &= \langle \psi^L(z)\psi^{\mathrm{new}}(0)\rangle, \\
G_{NN}(z) &= \langle \psi^{\mathrm{new}}(z)\psi^{\mathrm{new}}(0)\rangle.
\end{align}
Here $z$ is a complex coordinate (e.g. boundary coordinate), and analytic continuation corresponds to
\begin{equation}
z \;\longrightarrow\; e^{2\pi i} z.
\end{equation}

\paragraph{Monodromy constraints on correlators} We now impose consistency under analytic continuation. The key principle is that the correlation function after continuation must equal the correlation function of the transformed fields:
\begin{equation}
\langle \phi_1(e^{2\pi i}z)\phi_2(0)\rangle
=
\langle \mathcal{M}\phi_1(z)\,\mathcal{M}\phi_2(0)\rangle.
\end{equation}
We apply this to each correlator.

\paragraph{Left-left correlator}

Since $\psi^L$ is invariant under monodromy
\begin{equation}
\psi^L \to \psi^L,
\end{equation}
we obtain
\begin{equation}
G_{LL}(e^{2\pi i}z) = G_{LL}(z).
\end{equation}
Thus $G_{LL}$ must be single-valued.

\paragraph{Mixed correlator} For $G_{LN}$, we use
\begin{equation}
\psi^{\mathrm{new}} \to \psi^{\mathrm{new}} - 2\pi i\,\psi^L,
\end{equation}
so that
\begin{align}
G_{LN}(e^{2\pi i}z)
&= \langle \psi^L(z)\,(\psi^{\mathrm{new}}(0) - 2\pi i\,\psi^L(0)) \rangle \\
&= G_{LN}(z) - 2\pi i\, G_{LL}(z).
\end{align}

\paragraph{Log-log correlator} For $G_{NN}$, both insertions transform:
\begin{equation}
\psi^{\mathrm{new}} \to \psi^{\mathrm{new}} - 2\pi i\,\psi^L.
\end{equation}
Thus
\begin{align}
G_{NN}(e^{2\pi i}z)
&= \langle (\psi^{\mathrm{new}} - 2\pi i\,\psi^L)(z)\,(\psi^{\mathrm{new}} - 2\pi i\,\psi^L)(0) \rangle \\
&= G_{NN}(z)
- 2\pi i\, G_{LN}(z)
- 2\pi i\, G_{NL}(z)
+ (2\pi i)^2 G_{LL}(z).
\end{align}
Assuming symmetry of correlators ($G_{NL}=G_{LN}$), this simplifies to
\begin{equation}
G_{NN}(e^{2\pi i}z)
=
G_{NN}(z)
- 4\pi i\, G_{LN}(z)
+ (2\pi i)^2 G_{LL}(z).
\end{equation}

\paragraph{Solving the functional equations}

We now solve these constraints.

\paragraph{Form of $G_{LL}$} Single-valuedness implies that $G_{LL}$ has no logarithmic branch cuts. By scale invariance, the general form is
\begin{equation}
G_{LL}(z) = \frac{A}{z^{2h}}.
\end{equation}
Scale invariance and single-valuedness imply that the general form of the correlator is
$$
G_{L L}(z)=\frac{A}{z^{2 h}} .
$$
However, in the presence of a nontrivial Jordan block structure, a nonzero value of $A$ would correspond to a diagonalizable pairing between states, which is incompatible with the indecomposable action of the monodromy operator. We therefore set
$$
A=0, \quad \Rightarrow \quad G_{L L}(z)=0 .
$$

\paragraph{Solve for $G_{LN}$} The functional equation reduces to
\begin{equation}
G_{LN}(e^{2\pi i}z) = G_{LN}(z).
\end{equation}
Thus $G_{LN}$ is single-valued. Scale invariance then fixes
\begin{equation}
G_{LN}(z) = \frac{b}{z^{2h}}.
\end{equation}

\paragraph{Solve for $G_{NN}$} The equation becomes
\begin{equation}
G_{NN}(e^{2\pi i}z)
=
G_{NN}(z)
- 4\pi i\, \frac{b}{z^{2h}}.
\end{equation}
We now look for a function whose monodromy reproduces this shift. Recall that
\begin{equation}
\ln(e^{2\pi i} z) = \ln z + 2\pi i.
\end{equation}
Thus the function
\begin{equation}
G_{NN}(z) = \frac{\alpha \ln z + \beta}{z^{2h}}
\end{equation}
transforms as
\begin{align}
G_{NN}(e^{2\pi i}z)
&= \frac{\alpha (\ln z + 2\pi i) + \beta}{z^{2h}} \\
&= G_{NN}(z) + \frac{2\pi i\, \alpha}{z^{2h}}.
\end{align}
Matching with the required transformation gives
\begin{equation}
2\pi i\, \alpha = -4\pi i\, b \quad \Rightarrow \quad \alpha = -2b.
\end{equation}
Therefore,
\begin{equation}
G_{NN}(z) = \frac{-2b \ln z + c}{z^{2h}}.
\end{equation}

\paragraph{Final result}

Collecting the results, we obtain
\begin{align}
G_{LL}(z) &= 0, \\
G_{LN}(z) &= \frac{b}{z^{2h}}, \\
G_{NN}(z) &= \frac{-2b \ln z + c}{z^{2h}}.
\end{align}

\paragraph{Interpretation}
The logarithmic structure of the two-point functions is thus not imposed but follows directly from the monodromy representation. The appearance of the logarithm is required to reproduce the additive shift generated by the unipotent monodromy.

In particular, the coefficient of the logarithm is fixed entirely by the mixing term in the monodromy transformation, demonstrating that the Jordan block structure and the logarithmic behavior of correlators have a common geometric origin.

This provides a bulk derivation of logarithmic conformal field theory correlators, in which the logarithmic terms arise as a direct consequence of analytic continuation and monodromy, rather than being introduced as an independent assumption.

\section{Summary and outlook}
In this work, we have analyzed the logarithmic sector of critical topologically massive gravity (CTMG) at the chiral point $\mu \ell=1$ from the perspective of analytic continuation and monodromy. By constructing the logarithmic graviton as a derivative of the massive mode at the point of degeneracy, we showed that it develops a natural multivalued structure in the complexified radial coordinate, with logarithmic branch points arising at the zeros of $\cosh \rho$. This leads to a nontrivial monodromy representation acting on the space of linearized solutions, in which the left-moving and logarithmic modes form a rank-2 indecomposable (Jordan block) structure.

This monodromy naturally induces a decomposition of the solution space into untwisted and twisted sectors, distinguished by their transformation properties under analytic continuation around branch points. In this framework, the left-moving graviton belongs to the untwisted sector, while the logarithmic mode generates a twisted sector characterized by unipotent monodromy and non-diagonalizable action.

We further showed that these monodromy constraints are sufficiently restrictive to determine the characteristic logarithmic structure of two-point functions up to normalization, reproducing the expected form of logarithmic conformal field theory correlators. Importantly, this derivation arises purely from bulk analytic structure, without assuming logarithmic CFT data at the outset.

Overall, these results suggest that monodromy provides a natural organizing principle for logarithmic sectors in CTMG, unifying the appearance of Jordan block structures, branch point behavior, and logarithmic correlators within a single geometric framework. This perspective supports the interpretation of logarithmic modes as bulk analogues of twist-like fields and highlights a concrete link between gravitational dynamics, analytic continuation, and logarithmic conformal structures.

Several directions naturally follow from the monodromy-based perspective developed here. A first step would be to relate the monodromy representation of logarithmic modes more directly to the Chern-Simons formulation of three-dimensional gravity, where bulk geometries are encoded in flat $S L(2, \mathbb{R}) \times S L(2, \mathbb{R})$ connections. In this setting, it would be interesting to understand whether the unipotent Jordan structure identified here can be reinterpreted in terms of non-diagonalizable holonomies and how this description interfaces with the appearance of logarithmic sectors at the chiral point.

A second direction concerns the precise relation between the bulk branch structure and the combinatorial data of branched coverings. A natural next step is to establish a precise mapping between the branch points of $\ln \cosh \rho$ and the ramification data of Hurwitz theory. Clarifying this connection could provide a more direct geometric realization of the integrable structures and tau-function descriptions that appear in related approaches.

Finally, it would be of interest to extend the present analysis beyond the linearized level. In particular, understanding how monodromy and indecomposable structures manifest in higher-point functions or in the full non-linear theory could shed light on whether the logarithmic sector admits a consistent completion beyond perturbation theory. Such an extension may also clarify the role of twisted sectors in the full quantum theory of CTMG and their potential implications for the structure of the dual logarithmic conformal field theory.

%====================================================================
\paragraph{Acknowledgements} The author would like to thank Daniel Grumiller for his response to the author's query concerning the linearly growing contribution indexed by $\rho$ in the Fefferman-Graham expansion of the metric \cite{Grumiller:2008qz,Grumiller:2013at}. The author acknowledges financial support from the Department of Physics at the University of Pretoria.
%====================================================================

\clearpage

\bibliographystyle{utphys}
\bibliography{sample}
\end{document}